\documentclass[twocolumn,secnumarabic,amssymb, nobibnotes, aps, prd, showpacs,amsmath]{revtex4-1}
\usepackage[dvipdf]{graphicx}

\usepackage{graphicx}
\usepackage{dcolumn}
\usepackage{bm}
\usepackage{color}
\usepackage{cleveref}
\begin{document}
\title{Stabilizing topological states in a dynamic network of FitzHugh-Nagumo systems}
\author{Resmi V.}
\email[]{v.resmi@gmail.com}
\author{G. Ambika}
\email{g.ambika@iiserpune.ac.in}
\affiliation{Indian Institute of Science Education and Research, Pune-411 008, India}
\date{\today}
\begin{abstract}

We report how strategic evolution can stabilize topological states in a network of FitzHugh-Nagumo systems. The evolution follows a repeated process of adding or deleting of links between two nodes that is decided based on a threshold set for the closeness between them.  This results in a need based rewiring that keeps the systems in synchrony in a single cluster or in phase locked states in different clusters. We analyse in detail the occurrence of such multi stable topological fixed points with corresponding emergent dynamical states using the frequency of occurrence of each state when evolved from a large number of initial states. We find that there is an optimum range for the threshold used in the strategy that results in maximum frequency for the stable topological states. 
\end{abstract}

\pacs{05.45.Xt, 89.75.Fb, 87.18.Sn}

\maketitle
\section{Introduction}
Complex networks have emerged as an area of intense research activity in recent times. This is because they provide an effective framework for representing and modelling many types of phenomena occurring in complex systems. Such complex systems arise in several diverse areas, ranging from neuronal systems to social sciences \cite{Newmanbook1,Newmanbook2, Dorogovtsevbook, Estradabook}. In these cases, the evolution of the network and hence the emergence of collective behaviour depend both on the nodal dynamics and the network structure or the topology of the connectivity among the nodes. Earlier studies were mainly on static networks where the topology remains the same but dynamics at the nodes evolve under the topology giving rise to interesting dynamics like synchronization, clusters, amplitude death etc \cite{Bar02,Dor08,Hou03}.  
Recent reports indicate how rewiring mechanisms, where the topology is repeatedly changed, can enhance the richness of the dynamics \cite{ Hagberg08, Zeng12, Tessone12}. It is now well established that nodal dynamics and topology are equally important in real world situations and their interplay can lead to interesting results. A well defined network structure is often an important feature whose evolution can affect relevant biological processes. At the same time, the topology of connections obviously plays a role in information processing, spreading of diseases etc.  However most of the network growth methods reported, do not rely on nodal dynamics, but depends on structural changes alone to produce desired results \cite{Son12}. 

Several cases of synchrony and phase locking can be found in central nervous system and the resultant coherent activity plays an important role in behaviour and cognition. In this context, phase locked cluster oscillations are observed in periodically forced neuronal networks and also in networks with coupling of inhibitory and excitatory nature\cite{Langdon12}. Also it is reported how a fully synchronised single cluster breaks into two anti phase clusters due to phase resetting controls\cite{Achuthan09}. Co existent clustered states depending on initial conditions are shown to arise with inhibitory coupling\cite{Vreeswijk96}. Thus it is important to understand the mechanism responsible for the formation of phase locked clusters and other possible network structures and the role of nodal dynamics in stabilising them.

In this paper, we report how periodic attachment or removal of links based on a chosen strategy can establish coherent structures in a complex network. These emerge as the network evolves under both nodal dynamics and the proposed connectivity changes. We are motivated by the fact that real world networks often evolve with changes in topology in which links can appear or disappear as the whole system evolves. This is especially relevant in biological and social networks where contacts between systems are not constantly realised but occasionally activated \cite{Li11}. Such strategic evolution of topology is shown to result in favourable outcomes in social networks \cite{Bhat12}. It is reported that activity dependent interactions in a network of phase oscillators  can induce  coherent state, two cluster state, and chaotic state \cite{Aoki11}. It has been shown that in the context of FitzHugh-Nagumo (FHN) oscillators, rewiring the interaction pattern will induce collective firing \cite{Tessone12} such that lack of connectivity is compensated by dynamic rewiring \cite{Zan11,Kur95}. So also, adaptive coupling to enhance synchrony has been studied earlier \cite{Mei09,Bor03,Gro08,Grossbook,Win12}. The interplay between dynamics and topology is highly relevant in deriving dynamical criticality and optimal properties.  Along these lines, several evolutionary design algorithms have been introduced for coupled  phase oscillators \cite{Lev12,Mar09,Lou12,Zan05}. It is also reported how  
functional sub networks can be evolved according to the network structure and dynamics \cite{Li10}. These studies indicate that rewiring or dynamic evolution enhances the scope and applicability of networks and result in interesting and richer dynamical states.

In our method we follow a strategic growth of the network where connectivity changes are done periodically using a condition that is based on relative performance as governed by the dynamics at the nodes. This method, while changing the total links at every re wiring, finally drives the topology to stable states and thus provides a very simple mechanism for growing networks and stabilising them to different topologies. We illustrate this in networks with FHN neuronal model as nodal dynamics.  Starting from a very sparse network, we show how selective removal and adding of links based on this specific strategy can result in phase locked clusters in which nodes in each cluster are in almost synchrony, but are phase locked with nodes of the other cluster. It is interesting to see a stable bipartite structure emerging under such an evolution with two clusters anti phase to each other. Moreover the specific strategy chosen here leads to a structure in which there are no links among synchronised nodes in the same cluster but all to all connections between the nodes in different clusters.  We consider a large number of realisations with different starting initial conditions to establish that such an emergent topology is preferred by the system under the strategy used by us. We indicate how to establish the stability of the resultant topology, its dependence on the parameters in the design of the strategy and the role of nodal dynamics in this scheme.

\section{Evolving topology in networks}
We consider the FitzHugh Nagumo (FHN) neuron model as the nodal dynamics in our study. It is widely used to study the dynamics of neurons especially neuronal synchronization and desynchronization  phenomena in the brain. The intrinsic dynamics of FHN systems is modelled by the following equation,
\begin{eqnarray}
  \label{eq:fhn_single}
  \dot{x} &=&\frac{1}{\epsilon}( x - \frac{x^3}{3} - y ), \nonumber \\
  \dot{y} &=& a + x.
\end{eqnarray}
Here, $x_i$ represents the voltage of the neuron and $y_i$ represents the recovery variable. The variations in the voltage represent the spikes observed in neurons.  This system also can represent the dynamics of any excitable system since it has two time scales, which can be controlled by choosing appropriate values for the parameter $\epsilon$.

In the study reported in this paper, we consider a dynamic network of $N$ such FHN neurons coupled as 
\begin{eqnarray}
  \label{eq:fhn_net}
  \dot{x}_i &=&\frac{1}{\epsilon}( x_i - \frac{x_i^3}{3} - y_i ) + K \sum_{j=1}^{N} A_{ij} (x_j - x_i), \nonumber \\
  \dot{y}_i &=& a + x_i.
\end{eqnarray}
The strength of diffusive coupling is given by $K$. The network topology is decided by the elements the adjacency matrix $A$. If $A_{ij}=1$, the neurons $i$ and $j$ share a link. We choose the matrix $A$ to be symmetric  $A_{ij} = A_{ji}$ such that all couplings are bidirectional. Further, self-couplings are avoided by choosing $A_{ii}=0$. 

We introduce a strategy for evolving the system in which in addition to the dynamical evolution of the nodes, a new network topology is selected after a rewiring time of $\tau$. It is linked with nodal dynamics such that the strategy depends on the co-performance of nodes as indicated by the suitably defined distance between them in phase space. Thus the rewiring is based on local information on nodes and not on global measures like order parameter or total number of links etc.  To quantify this, we define the distance function between a pair of nodes, $\delta_{ij}$ as the Euclidean distance in phase space between them given by
\begin{equation}
  \delta_{ij} = \sqrt{(x_i-x_j)^2+(y_i-y_j)^2}
  \label{eq:delta}
\end{equation} 
(clearly $\delta_{ij}=\delta_{ji}$). 

The steps involved in the strategic evolution of the network are as follows.
\begin{enumerate}
\item  Start with a very sparse network of $N$ systems
\item After a rewiring time of $\tau$, check the value of $\delta_{ij}$  at very pair of nodes.
\item If  $ \delta_{ij} > \beta$, a chosen threshold, attach a link between $i$ and $j$ by setting $A_{ij}=1$. 
\item If  $ \delta_{ij} < \beta$, detach the link between $i$ and $j$ by setting $A_{ij}=0$.
\item Repeat steps $2$--$4$ after very time interval $\tau$.
\end{enumerate}

Thus in a way the matrix $A$ itself is also co evolving with the nodal dynamics. Here, the network is potentially global, which means that, any neuron can get connected to any other neuron, based only on the Euclidean distance between them. The evolution is continued till the network topology stabilises. In the next section, we present the detailed analysis on the topological states of the network that result from the  evolution scheme introduced. 

\section{Analysis}
The analysis is done numerically by using the parameters $a=0.95$, $\epsilon = 0.01$ such that the intrinsic dynamics is periodic spiking. The typical reset time is chosen to be $\tau=10$ while the coupling strength is varied from $K=0$ to $K=2$ and the threshold is varied from $\beta=0.1$ to $\beta=1$. For the numerical integration of Eq.~\ref{eq:fhn_net}, we use Adams-Bashforth-Moulton predictor-corrector method \cite{Sastrybook}.

\begin{figure}
 \centerline{\includegraphics[width=0.95\columnwidth]{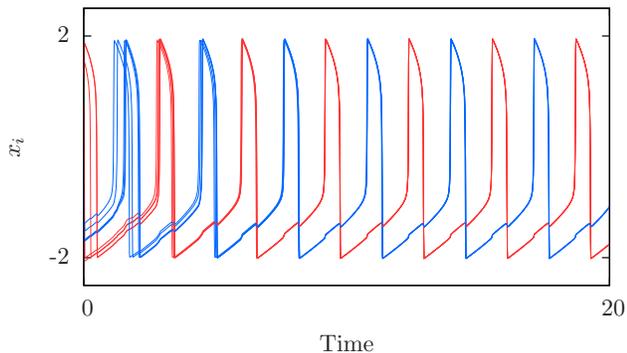}}
 \caption{ \label{fig:ts_10fhn_2clus}. Time series of the voltages $x_i$ of ten coupled FitzHugh-Nagumo systems ( Eq.~\ref{eq:fhn_net} ) showing $2$-cluster state for $K=1$, $\tau=10$ and $\beta=0.2$. In this case, both clusters have $5$ systems each. Here, the parameters of the individual systems are kept at $a=0.95$ and $\epsilon=0.01$.  }
\end{figure}

\begin{figure}
\centerline{\includegraphics[width=0.60\columnwidth]{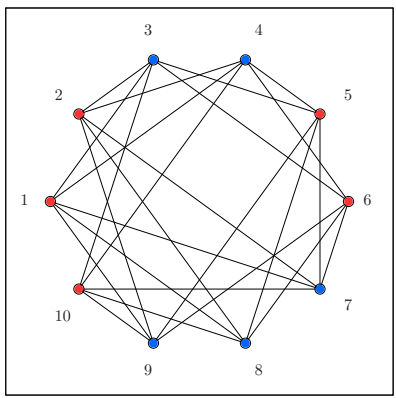}}
\caption{ \label{fig:bipart_example} A representation of the stable bipartite network. Blue and red nodes form two different groups (clusters). Here, both clusters have $5$ nodes each. Note that there is no two blue nodes ( and no two red nodes ) share a link and that all pairs of blue and red nodes are connected.  }
\end{figure}

\begin{figure}
\centerline{\includegraphics[width=0.60\columnwidth]{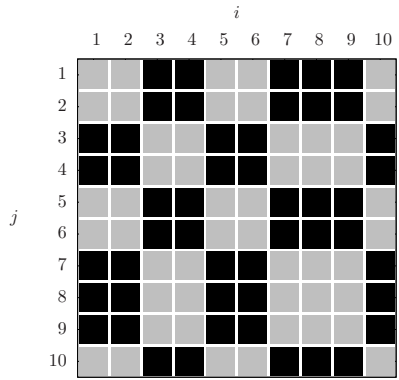}}
\caption{ \label{fig:bipart_example_adj} Adjacency matrix corresponding to the bipartite network shown in Fig.~\ref{fig:bipart_example}. Here, black squares represent ones and grey squares represent zeros. The adjacency matrix is symmetric since we consider all connections to be bidirectional. Note that rows $1$,$2$,$5$,$6$ and $10$ are similar and that rows $3$,$4$,$7$,$8$ and $9$ are similar. This corresponds to the cluster structure seen in Fig.~\ref{fig:bipart_example}.}
\end{figure}

We consider ten FHN neurons in random topology evolving under the strategy we have introduced. From numerical integration of Eq.~\ref{eq:fhn_net}  starting from random topology and random initial conditions, we find that, for certain initial conditions the systems stabilize to a $2$-clustered state for suitable values of coupling strength.  The time series from all the nodes for this   state is shown in Fig.~\ref{fig:ts_10fhn_2clus}. 
It is clear that five of the nodes are in synchrony and are in anti phase with the remaining five nodes. Thus this is a 2 cluster network with five nodes in each. As the network topology evolves with the system dynamics, it stabilizes to a bipartite structure with nodes in different clusters forming two groups. Due to the specific strategy chosen, there is no connection among the nodes in synchrony within a cluster while all pairs of nodes from different clusters are connected. The resulting network structure is shown in Fig.~\ref{fig:bipart_example}.
The corresponding adjacency matrix is shown in Fig.~\ref{fig:bipart_example_adj}. Here, black squares represent ones and grey squares represent zeros. The adjacency matrix is symmetric since we consider all connections to be bidirectional.

We proceed with the numerical study by choosing large number of different initial conditions and find that, different types of bipartite networks ($5$ \& $5$, $4$ \& $6$, $3$ \& $7$, $2$ \& $8$ and $1$ \& $9$ ) are possible for different sets of initial conditions in phase space. We use the term $2$-cluster state to refer to all these states collectively. There are also some initial conditions that evolve into a single cluster or a $1$-cluster state, with all the systems in almost synchrony and the resulting topology is a set of disconnected nodes. By almost synchrony we mean, as mentioned in \cite{Achuthan09} the difference in the firing times of two systems is less than ten $\%$ of their inter spike interval. These different topological states therefore multi stable states of the system.

To identify all the possible resultant network topologies, we first find the number of clusters( number of clusters for a bipartite network is $2$, for a tri-partite network is $3$  etc.) from the adjacency matrix as described below. Then we find the number of systems in each cluster from the total links in the evolved topology. As mentioned above, the systems which are synchronized among each other are not connected and the corresponding elements in the adjacency matrix are zeroes ($A_{ij}=A_{ji}=0$). Conversely, if $A_{ij}=1$ then, systems at nodes $i$ and $j$ are not synchronized and thus, belong to different clusters. This means that, the number of clusters in each topology can be identified as the number of unique rows in the adjacency matrix. For example, in the adjacency matrix shown in Fig.~\ref{fig:bipart_example_adj}, we find that there are only two unique rows (or columns, since the matrix is symmetric) namely, row-$1$ and row-$3$. All other rows are replicas of these two unique rows. Thus, we identify the network represented by the adjacency matrix shown in Fig.~\ref{fig:bipart_example_adj} as bipartite. The $1$-cluster can also be identified as the network with number of links, $M=0$. 

As noted earlier, all pairs of nodes in different clusters are connected to each other. Thus, the number of links in a $2$-clustered state of ($1$ \& $9$) can be estimated to be $9$, since the only node in the first cluster is connected to all $9$ nodes in the second cluster, and no two nodes in second cluster share a link. Similarly, ($2$ \& $8$) clustered state has $M=16$ links, ($3$ \& $7$) clustered state has $M=21$ links, ($4$ \& $6$) clustered state has $M=24$ links, and ($5$ \& $5$) clustered state has $M=25$ links. In the case of $3$-clustered states the number of unique rows in the adjacency matrix is found to be $3$. The number of links in a $3$-cluster state of ($1$, $1$ \& $8$) can be estimated to be $M=17$, since the only nodes in first two clusters make one link with each other and $8$ nodes each with the third cluster. Similarly, ($1$, $2$ \& $7$) clustered state has $M=23$ links, ($1$,$3$ \& $6$) clustered state has $M=27$ links, ($1$, $4$ \& $5$) clustered state has $M=29$ links, ($2$, $2$ \& $6$) clustered state has $M=28$ links, ($2$, $3$ \& $5$) clustered state has $M=31$ links, ($2$, $4$ \& $4$) clustered state has $M=32$ links and ($3$, $3$ \& $4$) clustered state has $M=33$ links. Thus, each resultant topology can be identified from the number of unique rows in the adjacency matrix and the number of links.

\section{Topological fixed points}
\begin{figure}
\centerline{\includegraphics[width=0.95\columnwidth]{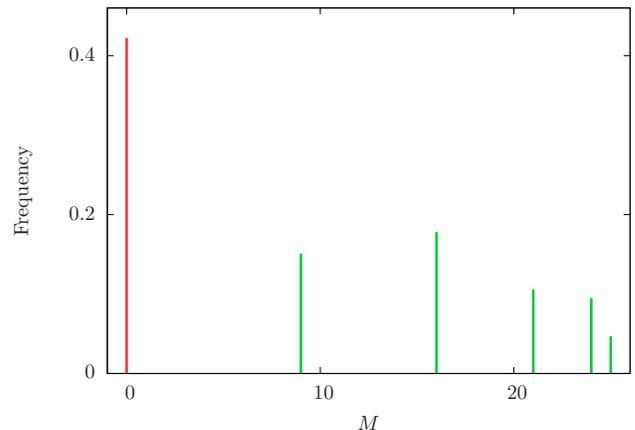}}
\caption{ \label{fig:hist_ep_2}. Frequency of occurrence of each network topology for a net work of ten FHN systems. The parameters are chosen to be $K=2.0$, $\beta=0.2$ and $\tau=10$. Red line at $M=0$ corresponds to $1$-cluster state with $0$ links. Green lines correspond to different $2$-cluster topologies. For details, see text. }
\end{figure}

\begin{figure}
  \centerline{\includegraphics[width=0.95\columnwidth]{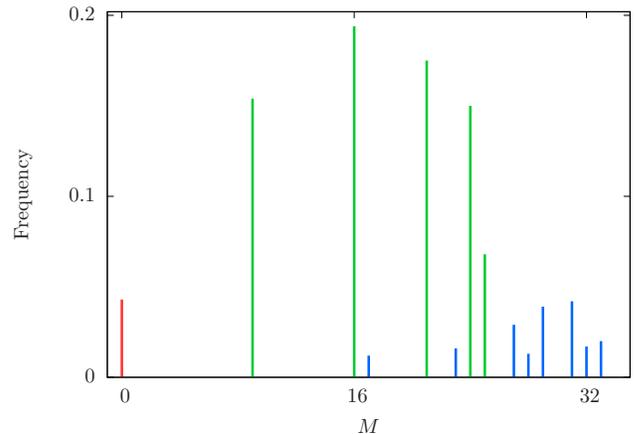}}
  \caption{ \label{fig:hist_ep_p2}. Frequency of occurrence for each network topology for $N = 10$. The parameters are chosen to be $K=0.2$, $\beta=0.2$ and $\tau=10$. Red line at $M=0$ corresponds to $1$-cluster state with $0$ links. Green lines correspond to different $2$-cluster topologies and blue lines to various $3$ cluster states. For details of various possible sub types in $2$ and $3$ clusters, see text. }
\end{figure}

\begin{figure}
\centerline{\includegraphics[width=0.90\columnwidth]{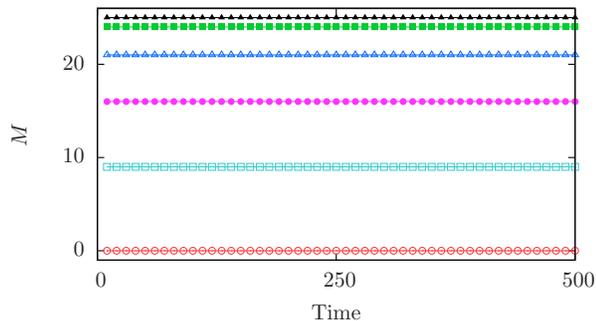}}
\caption{\label{fig:ts_M} A representative figure showing the topological fixed points. Here the number of links $M$ is plotted against time for a network of ten FHN neurons at coupling strength $K=1$ and reset time $\tau=10$. The value of $\beta$ is chosen to be $0.5$. Note that the time is shown in units of reset time, $\tau$. }
\end{figure}

As we have noted earlier, we find our method results in multi stable topological states depending on their initial states. It is to be noted that by initial state we mean both the starting topology and the initial conditions in phase space for the nodes. The number of systems in each cluster or the variety of possible clusters also depends on initial starting states. To establish the multi stability and to get a comprehensive picture of stabilised topologies, we consider $1000$ realisations of the nodal states and evolve them based on the method described in the previous section. For each stable $1$-cluster and $2$-cluster topology, we calculate the frequency of occurrence defined as, 
Frequency = number of occurrences of a certain topology/Total number of initial conditions. 
This can be understood as the probability of occurrence of each stable topology. This result is used to plot the histogram shown in Fig.~\ref{fig:hist_ep_2}. The parameters are chosen as $K=2.0$, $\beta=0.2$ and $\tau=10$. Here the red line at $M=0$ corresponds to $1$-cluster state with $0$ links while the green lines correspond to different $2$-cluster topologies. We find that the $1$-cluster state has the maximum frequency of occurrence for this set of parameters, while the $3$-cluster state does not stabilise. We identify the various $2$ cluster states from the number of links $M$. Thus green line at $M=9$ corresponds to bipartite cluster($1$,$9$), that at $M=16$ corresponds to ($2$, $8$), at $M=21$ to ($3$, $7$), at $M=24$ to ($4$, $6$) and at $M=25$ to ($5$, $5$) cluster.
We repeat the analysis for a smaller value of the coupling strength and see that, then $3$-cluster states are also present in addition to $1$-cluster and $2$-cluster states. The dynamics in these clusters is phase locked with each other. A histogram showing the frequency of occurrence of $1$,$2$ and $3$ clustered state for $K=0.2$ is given in Fig.~\ref{fig:hist_ep_p2}. Here we choose red line at $M=0$ to represent  $1$-cluster state with $0$ links. The green lines in the figure correspond to different $2$-cluster topologies and blue lines to various $3$ cluster states. We find that, the $2$-cluster state with  $2$ \& $8$ nodes has the maximum frequency of occurrence for this set of parameters. Green line at $M=9$ corresponds to ($1$ \& $9$) cluster.  The various possible clusters within the $2$ cluster and $3$ cluster topologies can be identified in the figure as follows: Green lines at $M=16$ corresponds to ($2$ \& $8$) cluster.  Green line at $M=21$ corresponds to ($3$ \& $7$) cluster.  Green line at $M=24$ corresponds to ($4$ \& $6$) cluster. Green line at $M=25$ corresponds to ($5$ \& $5$) cluster. Blue line at $M=17$ corresponds to ($1$, $1$ \& $8$) cluster, at $M=23$ corresponds to ($1$, $2$ \& $7$) cluster, at $M=27$ corresponds to ($1$,$3$ \& $6$) cluster, at $M=28$ corresponds to ($2$, $2$ \& $6$) cluster, $M=29$ corresponds to ($1$, $4$ \& $5$) cluster, at $M=31$ corresponds to ($2$, $3$ \& $5$) cluster, at $M=32$ corresponds to ($2$, $4$ \& $4$) cluster and at $M=33$ corresponds to ($3$, $3$ \& $4$) cluster.
Thus we see that the network settles to different final topologies depending on parameter values, coupling strength and initial conditions. We refer to these as topological fixed points of the system since the network settles to a final topology which does not change further with time. To identify stable fixed-point topologies, we evolve the coupled system from $1000$ different initial conditions, for $500 \times \tau$ time steps. The data for the first $300 \times \tau$ time steps is neglected as transients. We take the time-average and the standard deviation of $M$ for each realization. If the standard deviation, $sd_{M} < 0.1 $, we consider the topology to be a topological fixed point. We mention that the fixed point nature is in the topology and not in the nodal dynamics since even in the final state, the nodes in the topological fixed point will have dynamics as given by Eq.~\ref{eq:fhn_net}. The number of links in the different fixed point topologies remains constant in time. This is clear from Fig.~\ref{fig:ts_M}, where we plot the number of links at different reset times.

\begin{figure}
\centerline{\includegraphics[width=0.90\columnwidth]{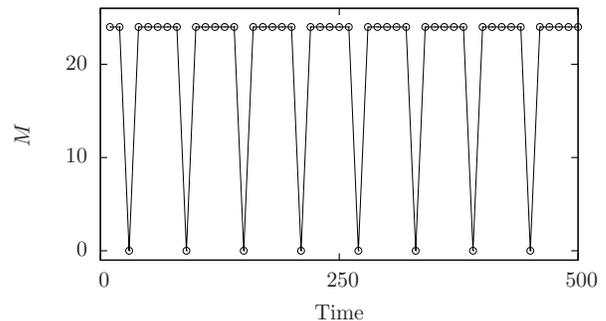}}
\caption{\label{fig:ts_M_switch} Number of links $M$ against time for a network of ten FHN neurons for $\beta=1$. The other parameters are kept fixed at the values given in Fig.~\ref{fig:ts_M}. Here also, the time is measured in units of reset time, $\tau$. We see that the topology switches between a $1$-cluster state of no connections ($M=0$) and a bipartite state ($M=24$).}
\end{figure}

However we find that while most of the $1000$ initial conditions get stabilized to a $1$-cluster or $2$-cluster state, there are a few remaining initial conditions that do not stabilise to any fixed point topology. Among them, especially for larger values of $\beta$, we find an interesting behaviour of switching between topological states. This is illustrated in Fig.~\ref{fig:ts_M_switch} by plotting the number of links as a time series against the rest times for $\beta=1$. 

To further establish the stability of the observed topological fixed points, we start the numerical integration of Eq.~\ref{eq:fhn_net} from a random network of ten FHN neurons for the set of parameter values $K=2$, $\beta=0.5$ and $\tau=10$. After $5000$ time steps ($500$ resets), we find that the system gets settled to a topological state, either a $1$-cluster or $2$-cluster state. To test the stability of this resultant topology, we disturb one link of the resultant topology ie. we choose a link at random and if the element corresponding to that link in the adjacency matrix is one, we change it to zero and vice-versa. Then the system is evolved for some more time and we see that, the system goes back to its original topology in one or two reset times. We repeat the analysis for a different values of  $\tau=1$ to confirm that the topology is stable against random perturbations.

The results presented so far are done considering a random initial topology. To study whether the initial topology affects the results, we repeat the numerical analysis with different starting topologies. For example, we start from $5$ \& $5$ bipartite cluster with $K \leq 0.4$ and $\beta=0.1$ and get single cluster ($M=0$) and various $2$-clusters and $3$-clusters as final topologies. This is repeated with all-to-all connected topology and $1$-clustered ($M=0$) topology as initial states,  and still they end up in multi stable topological states. 
\section{Role of nodal dynamics and parameters}
To check whether the results seen are the effects of the intrinsic dynamics of the systems or the strategy, we consider ten FitzHugh-Nagumo systems in all-to-all coupled topology and let them evolve for $5000$ time units with out applying the strategy discussed above. We find that, for sufficiently large coupling strengths ($K \sim 1$), depending on initial conditions, synchronized states where all systems are in one cluster and clustered states where systems divide themselves into two clusters can happen. We find qualitatively similar results for the simulations starting with random network topologies. Though the two clustered state coexisting with the fully synchronized state seems counter intuitive similar results has been reported in the case of two coupled Lorenz systems \cite{Cam12}. 

\begin{figure}
  \centerline{\includegraphics[width=0.95\columnwidth]{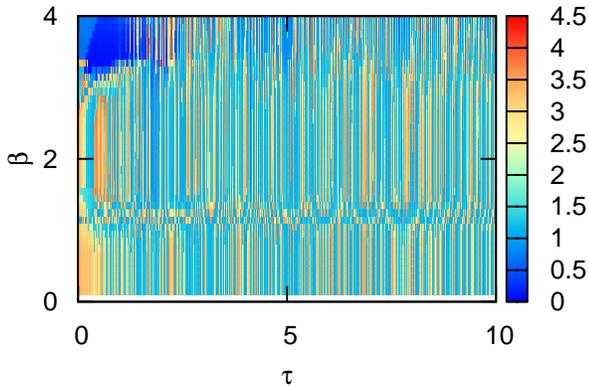}}
  \caption{\label{fig:pp_random} Regions of synchronization in the parameter plane $\tau$--$\beta$ for two strategically coupled neurons. For each pair $\tau$ and $\beta$, we plot the Euclidean distance $\delta_{12}$ between two neurons as colour coded plot.}
\end{figure}

We next consider the limiting case of two systems. In this case also, depending on initial conditions, the two systems can evolve into in-phase or anti-phase states. We further study the evolution of these two systems in the presence of the strategy discussed above. The parameter plane $\tau$--$\beta$ showing regions of synchronization is shown in Fig.~\ref{fig:pp_random}. In this case, we increase the value of $\tau$ from $0.01$ to $10$ in steps of $0.01$ and $\beta$ from $0.1$ to $4$ in steps of $0.1$. For each pair $\tau$ and $\beta$, we plot the Euclidean distance $\delta_{12}$ between two neurons as colour coded plot. Blue regions ( $\delta_{12} \sim 0$ ) thus correspond to synchronized states. The coupling strength is kept fixed at $K=1$. Initial conditions are $x_{1}=-1.779796$,$y_{1}=-0.820021$, $x_{2} = -1.965043$ and $y_{2}=0.527263$. We have verified that, starting from this set of initial conditions, the systems do not get synchronized in the absence of strategic evolution. 
This means that, by properly choosing the parameters $\beta$ and $\tau$, the systems can be driven from one topology of anti phase state to the other of in phase state. Hence, we infer that both the intrinsic dynamics and the strategic evolution have a role in stabilizing the topological states of the system. 

\begin{figure}
\centerline{\includegraphics[width=0.60\columnwidth, angle=90]{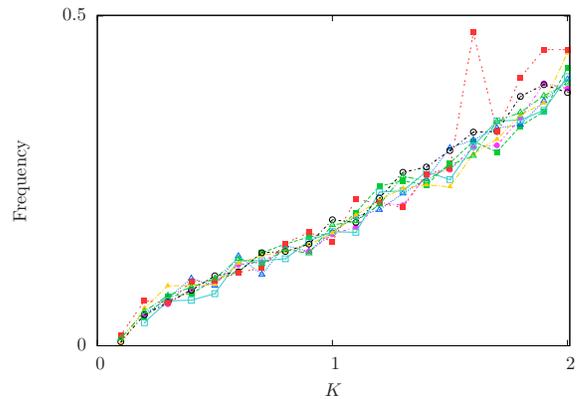}}
\caption{ \label{fig:1clus_freq_vs_K}. Frequency of occurrence of $1$-cluster state for increasing values of coupling strength. Different curves correspond to different values of $\beta$. In all cases, the frequency increases as $K$ is increased. }
\end{figure}

\begin{figure}
\centerline{\includegraphics[width=0.60\columnwidth,angle=90]{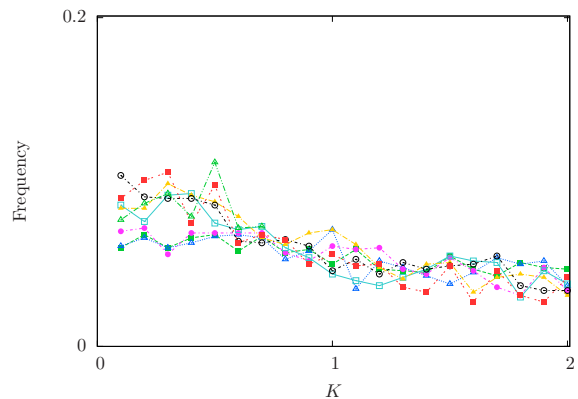}}
\caption{ \label{fig:2clus_5} Frequency of occurrence of $5$, $5$ cluster for increasing $K$. Curves of different colours correspond to different values of $\beta$. The frequency decreases for higher values of coupling strength. For $\beta=1.0$, this cluster occurs only for  $K \leq 0.7$. }
\end{figure}

We investigate further the role of the most relevant parameters like $K$, $\beta$ and $\tau$ in the outcome of stabilised topology. For this we do the full analysis for different values of $K$ and $\beta$ by keeping $\tau=10$. We find that in general the relative frequency of occurrence for each topology varies for different values of parameters. The change in frequency of $1$-clustered state for increasing $K$ is shown in Fig.~\ref{fig:1clus_freq_vs_K}. We see that for all curves, the frequency of occurrence increases as $K$ is increased.
Similarly, we illustrate this for the $2$ cluster state for the specific case of $5$, $5$ cluster in Fig.~\ref{fig:2clus_5}. Similar curves are obtained for other $2$-clustered states. We infer that as $K$ increases, the frequency of occurrence of $2$ cluster state decreases.

\begin{figure}
\centerline{\includegraphics[width=0.60\columnwidth,angle=90]{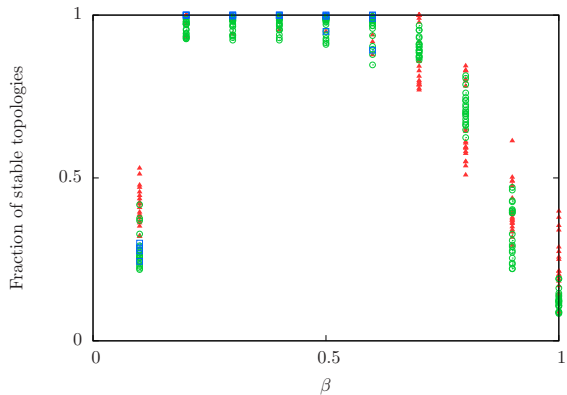}}
\caption{ \label{fig:stable_vs_thr} Fraction of stable fixed point topologies for increasing values of $\beta$. Here, we have considered $N=10$ neurons. Red triangles represent $\tau=10$, green circles represent $\tau=1$ and blue squares represent $\tau=0.01$. For the $\tau=0.01$ case, the analysis is done only for $\beta \leq 0.6$. Different points of the same colour represent different coupling strengths in the range $0.1$--$2$ increased in steps of $0.1$ for both $\tau=10$ and $\tau=1$ cases. For $\tau=0.01$ case, the coupling strength is increased in steps of $0.5$ in the range $0.5$--$2$.}
\end{figure}

We study the effect of varying $\beta$ on stabilisation of topologies. For this we compute the fraction of all the stable fixed point topologies for different values of $\beta$. In Fig.~\ref{fig:stable_vs_thr}, we plot this fraction for increasing values of $\beta$.  We see that, for very small value of $\beta$ ($\beta = 0.1$), more than half of the initial conditions leads to unstable topologies. As $\beta$ is increased, the fraction of stable topologies is found to increase.  We repeat the analysis for different values of coupling strength in the range $K=0.1$--$K=2$ and different values of reset times ($\tau=10$, $\tau=1$ and $\tau=0.01$) and the results are incorporated in Fig.~\ref{fig:stable_vs_thr}. 
It is interesting to note that, there is an optimal range of $\beta$, from $0.2$ to $0.7$, where most of the initial conditions lead to topological fixed points as the frequency of stable topologies reaches almost $1$. 
\section{Conclusions}
Dynamic neuronal networks form an important paradigm of increasing importance in brain research, especially related to the functional analysis of biological neuronal networks.  Often in such contexts it is important to study the temporal dynamics in relation to spatial patterns. Here we present a very simple evolutionary strategy that can utilise the mutual effect of spatial and temporal factors to stabilise a particular topology in the network.  The dynamic evolution studied is based on a rewiring scheme derived from local information on the co-performance of adjacent nodes. We find this facilitates the formation of bipartite structure with two phase locked clusters in addition to a prevalent single cluster.  We analyse the probability of occurrence of different possible bipartite topologies in a given network for specific chosen values of the parameters.  Considering the relevant parameters involved in the process as $\beta$, $\tau$ and $K$, we study in detail the effect of varying them on the final stabilised topology. Some of the important results and conclusions are given below.

For large value of $K$, the single cluster is mostly preferred, while smaller values of $K$ lead to bipartite and even tripartite topologies. For large value of the threshold $\beta$, switching between topologies occur. There is an optimal range of $\beta$ in which almost all initial conditions lead to topological fixed points. The dynamics at the nodes has a role to play in the evolution of fixed topology. In a network of same systems without any strategy also phase locked clusters are possible. What the strategy does is to drive the network to a stable topology and keep it there without any links within a cluster. In the limiting case of two systems, we find the strategy can change anti-phase to in-phase and vice versa. This means strategy can induce a tendency to settle to definite topologies. Conceptually, we feel the topology to which the network stabilises is similar to stable fixed point in dynamical systems. Accordingly, we establish the stability by considering small perturbations from the stable or fixed point topology. Even though multi stability for dynamical states have been reported for networks and other coupled systems, to the best of our knowledge, this is the first time multi stability in topology is studied in detail.

The homogeneous model of network studied to establish the growth of phase locked clusters may not capture all the complexities of neuronal dynamics or real world networks. However, it opens up the possibility of keeping the emergent state of a network in synchronization without any instantaneous links between their nodes but by keeping a strategy alive. Also the state is really robust since the moment small perturbations or noise arise due to external effects etc. the strategy will become active and change topology accordingly to retain the synchronization.  This need based re wiring can be optimised for minimum energy expenditure in engineering systems or real world networks with possible biological justifications.

Emergence of bipartite structure is interesting and is concurrent with the presence of motifs with fan like structure in the network. Usually bi partite networks are considered in relation to two different types of nodes. Here we show how bipartite nature emerges by dynamic evolution from a single type of nodes.  This can be considered as an extreme case of the conventional bi partites. 

It is well established that self organised structures can emerge from a collection of interacting elements. The study reported here confirms that  a strategy based dynamic evolution can drive the system to definite structures and stabilised topologies. Thus this provides one more instance that the co-evolution of dynamics and topology can naturally determine the network properties. We are aware that the procedure reported here needs further extensions to make it strictly applicable to real world situations. One way to do this is to introduce a probability in making and breaking of connections.  The work related to these issues is already in progress and will be reported elsewhere.
\section*{Acknowledgement}
We acknowledge DST, New Delhi, India for support through project SR/S2/HEP-08/2009.

\end{document}